\documentclass[12pt,twoside]{article}
\usepackage{fleqn,espcrc1,epsfig}
\newcommand{\bea}{\begin{eqnarray}}
\newcommand{\eea}{\end{eqnarray}}
\newcommand{\be}{\begin{equation}}
\newcommand{\ee}{\end{equation}}
\newcommand{\bt}{\begin{tabular}}
\newcommand{\et}{\end{tabular}}

\newcommand{\no}{\nonumber}

\newcommand{\beas}{\begin{eqnarray*}}
\newcommand{\eeas}{\end{eqnarray*}}

\newcommand{\AmS}{{\protect\the\textfont2
  A\kern-.1667em\lower.5ex\hbox{M}\kern-.125emS}}
\title{\vspace*{-1.7cm}Chiral Model for Dense, Hot and Strange Hadronic Matter 
\thanks{This work was funded in part by Deutsche 
Forschungsgemeinschaft (DFG)
and 
Bundesministerium f\"ur Bildung und Forschung (BMBF).}}
\author{D. Zschiesche\address{Institut f\"ur Theoretische Physik,
 D-60054 Frankfurt am Main, Germany},
        P. Papazoglou${}^{\rm{a}}$,
        Ch. W. Beckmann${}^{\rm{a}}$, 
        S. Schramm\address{GSI Darmstadt, Postfach 11 05 52, D-64220
Darmstadt, Germany}, 
J. Schaffner-Bielich
\address{Riken BNL Research Center, Brookhaven National Lab, Upton,
New York 11973}, H. St\"ocker${}^{\rm{a}}$, and W. Greiner${}^{\rm{a}}$}
\date{\today}
\begin{document}
\maketitle
\vspace{-0.5cm}
\section{Introduction}
Until now it is not possible to determine the equation of state (EOS)
of hadronic matter from QCD.  
One succesfully applied alternative way to describe the hadronic world at 
high densities and temperatures are effective
models like the RMF-models \cite{sero86}, where the relevant degrees of 
freedom are baryons and mesons
instead of quarks and gluons. Since 
approximate chiral symmetry is an essential feature of QCD, it
should be a useful concept for building and restricting effective models.
It has been shown \cite{furn93,heid94} that effective $\sigma-\omega$-models 
including SU(2) chiral symmetry are able to obtain a reasonable 
description of nuclear matter
and finite nuclei. 
Recently  \cite{paper3} we have shown that an extended 
$SU(3) \times SU(3)$ chiral ${\sigma-\omega}$ model is able to
describe nuclear matter ground state properties, vacuum properties and
finite nuclei satisfactorily. This model includes the lowest 
SU(3) multiplets of the baryons (octet and decuplet\cite{paper5}),
 the spin-0 and the spin-1 mesons
as the relevant degrees of freedom.
 Here we will discuss the predictions of this model for dense, hot, 
and strange hadronic matter.
\section{Nonlinear chiral SU(3) model}
\label{model}
We consider a relativistic field theoretical model of 
baryons and mesons built on
chiral symmetry and scale invariance. The general form of the
Lagrangean looks as follows:
\be
\label{lagrange}
{\cal L} = {\cal L}_{\mathrm{kin}}+\sum_{W=X,Y,V,{\cal A},u}{\cal L}_{\mathrm{BW}}+
{\cal L}_{\mathrm{VP}}
+{\cal L}_{\mathrm{vec}}+{\cal L}_{0}+{\cal L}_{\mathrm{SB}} .\no
\ee
${\cal L}_{\mathrm{kin}}$ is 
the kinetic energy term, ${\cal L}_{\mathrm{BW}}$ includes the  
interaction terms of the different baryons with the various spin-0 and spin-1 
mesons. The baryon masses are generated by the nonstrange
(${<q\bar{q}>}$)  
scalar condensate
$\sigma$ and the strange (${<s\bar{s}>}$) scalar condensate
$\zeta$. ${\cal L}_{\rm{VP}}$ contains the interaction terms 
of vector mesons with pseudoscalar mesons. 
${\cal L}_{\rm{vec}}$ generates the masses of the spin-1 mesons through 
interactions with spin-0 mesons, and ${\cal L}_{0}$ gives the meson-meson 
interaction terms which induce the spontaneous breaking of chiral symmetry.
It also includes the scale breaking logarithmic potential. Finally, 
${\cal L}_{\mathrm{SB}}$ introduces an explicit symmetry breaking of the
U(1)$_A$, the SU(3)$_V$, and the chiral symmetry. 
All these terms have been discussed in detail in \cite{paper3}.\\
\label{mfa}
To ingvestigate hadronic matter
properties at finite density and temperature  we 
adopt the mean-field approximation, i.e. the meson field operators are
replaced by their expectation values and the fermions 
are treated as quantum mechanical one-particle operators \cite{serot97}. 
After performing these approximations, the Lagrangean (\ref{lagrange}) 
becomes
\begin{eqnarray*}
{\cal L}_{BM}+{\cal L}_{BV} &=& -\sum_{i} \overline{\psi_{i}}[g_{i 
\omega}\gamma_0 \omega^0 
+g_{i \phi}\gamma_0 \phi^0 +m_i^{\ast} ]\psi_{i} \\ \no
{\cal L}_{vec} &=& \frac{ 1 }{ 2 } m_{\omega}^{2}\frac{\chi^2}{\chi_0^2}\omega^
2  
 + \frac{ 1 }{ 2 }  m_{\phi}^{2}\frac{\chi^2}{\chi_0^2} \phi^2
+ g_4^4 (\omega^4 + 2 \phi^4)\\
{\cal V}_0 &=& \frac{ 1 }{ 2 } k_0 \chi^2 (\sigma^2+\zeta^2) 
- k_1 (\sigma^2+\zeta^2)^2 
     - k_2 ( \frac{ \sigma^4}{ 2 } + \zeta^4) 
     - k_3 \chi \sigma^2 \zeta \\ 
&+& k_4 \chi^4 + \frac{1}{4}\chi^4 \ln \frac{ \chi^4 }{ \chi_0^4}
 -\frac{\delta}{3}\ln \frac{\sigma^2\zeta}{\sigma_0^2 \zeta_0} \\ \no
{\cal V}_{SB} &=& \left(\frac{\chi}{\chi_0}\right)^{2}\left[m_{\pi}^2 f_{\pi} 
\sigma 
+ (\sqrt{2}m_K^2 f_K - \frac{ 1 }{ \sqrt{2} } m_{\pi}^2 f_{\pi})\zeta 
\right] , 
\end{eqnarray*}
with $m_i$ the effective mass of the baryon $i$  
($i=N,\Lambda,\Sigma,\Xi,\Delta,\Sigma^\ast,\Xi^\ast,\Omega$). 
$\sigma$
and $\zeta$ correspond to  
the scalar condenstates, $\omega$ and $\phi$ represent
the non-strange and the strange vector field respectively and $\chi$ is
the dilaton field.
Now it is straightforward to write down the expression 
for the thermodynamical potential of the grand canonical 
ensemble $\Omega$ per volume $V$ 
at a given chemical potential $\mu$ and temperature $T$
\be
   \frac{\Omega}{V}= -{\cal L}_{vec} - {\cal L}_0 - {\cal L}_{SB}
-{\cal V}_{vac}- \sum_i \frac{\gamma_i }{(2 \pi)^3}  
\int d^3k \left[\ln{\left(1+e^{-\frac 1T[E^{\ast}_i(k)-\mu^{\ast}_i]}\right)}
+  \ln{\left(1+e^{-\frac 1T[E^{\ast}_i(k)+\mu^{\ast}_i]}\right)}
\right] ,
\ee 
from which all thermodynamic quantites can be derived.
 $\gamma_i$ denote the fermionic 
spin-isospin degeneracy factors.
The single particle energies are 
$E^{\ast}_i (k) = \sqrt{ k_i^2+{m_i^*}^2}$ 
and the effective chemical potentials read
 $\mu^{\ast}_i = \mu_i-g_{i\omega} \omega-g_{i\phi} \phi$. 
The mesonic fields are determined by minimizing ${\Omega/V}$.
\section{Excited hadronic matter}
\begin{figure}[h]
\epsfig{figure=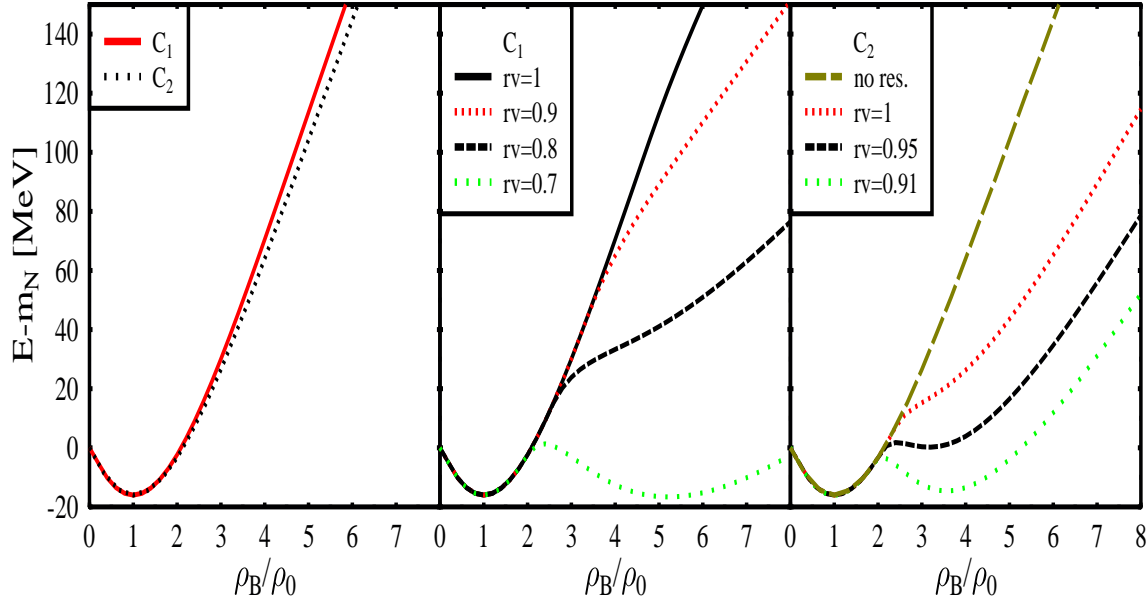,width=16cm,height=9cm}
\vspace*{-2cm}
\caption{\small  Nuclear matter equation of state 
for the parameter sets $C_1$ and $C_2$. (left: no baryon resonances, 
middle: $C_1$ including baryon resonances for different values of the quotient 
$r_v=\frac{g_{N\omega}}{g_{\Delta\omega}}$ right: same for $C_2$). 
\label{eos}}
\vspace*{-0.8cm}
\end{figure}

\begin{figure}[hb]
\centerline{\parbox[b]{10cm}{\epsfxsize=12cm
\vspace*{-1.5cm}
\epsfbox{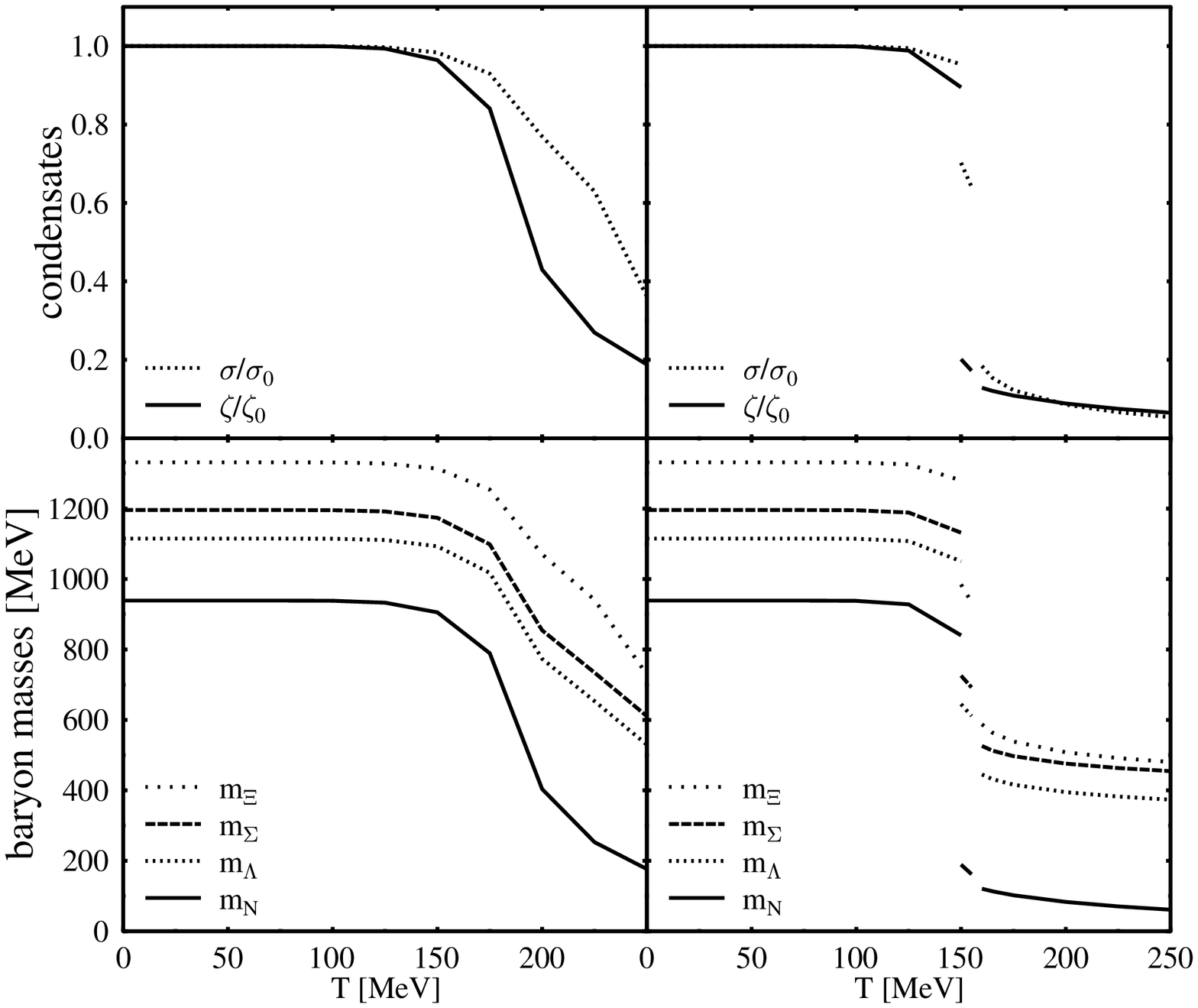}}\hfill
\parbox[b]{4.3cm}{\caption{\small Chiral condensates (top) and
resulting baryon masses (bottom) as a function
of temperature for vanishing chemical potential (Left: no resonances
included, right: baryon decuplet included).}\vspace{5cm}\label{condensates}}}
\vspace*{-1cm}
\end{figure}

\begin{figure}[ht]
\vspace{-1cm}
\centerline{\parbox[b]{7cm}{\epsfxsize=9.5cm
\epsfbox{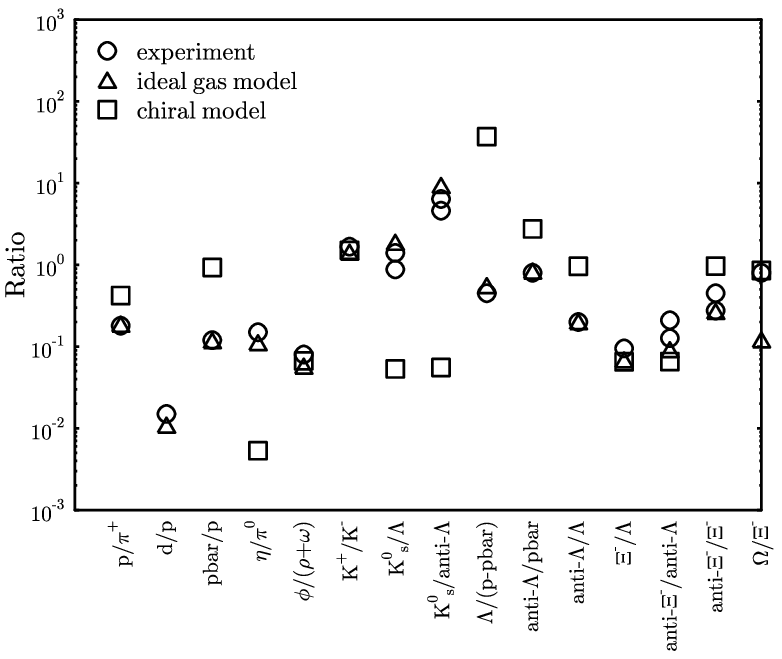}}\hfill
\parbox[b]{6.3cm}{\caption{\small} Particle ratios for 200 A GeV/c
S+Au collisions. Experimental yields are compared to ideal gas
and chiral model calculations 
 using $T=160 MeV$, $\mu_q=57 MeV$ and $\mu_s= 20 MeV$, 
as obtained from ideal gas fits. \vspace{3cm}\label{ratios}}}
\vspace{-1.5cm}
\end{figure}

The parameters are fixed by hadronic vacuum masses and nuclear
matter ground state properties \cite{paper3,paper5}.
For the following investigations we consider the 
two parameter sets $C_1$ and $C_2$, which satisfactorily describe finite 
nuclei \cite{paper3}. The main difference between the two parameter
sets is the 
coupling of the strange condensate to the nucleon and the $\Delta$. 
While in $C_2$ this coupling is set to zero, in the case of $C_1$ 
the nucleon and the $\Delta$ couple to
the $\zeta$ field. 
The equation of state for dense hadronic matter at vanishing
strangeness without baryon resonances is shown in
fig.\ref{eos}(left). In the middle and right picture the EOS
including resonances is plotted for a varying ratio
$r_v=g_{\Delta\omega}/g_{N\omega}$. 
One observes that for ${r_v=1}$ the resulting EOS strongly depends
on the way of the nucleon and $\Delta$ mass generation.
For a pure $\sigma$-dependence of the masses of the nonstrange baryons
($C_2$) the equation of state is strongly influenced at high densities
by the production of resonances, in contrast to the model
where both masses are partly generated by the strange condensates ($C_1$).
This is due to the different behaviour of the ratio of the effective masses
$m_\Delta^\ast/m_N^\ast$ as a function of density.
Furthermore one observes that the softening of the equation of state
at high densities strongly depends on the vector coupling of the resonances. 
If we assume that there are no $\Delta$'s in the
groundstate of nuclear matter and that density isomers are not
absolutely stable, the minimal value for the vector
coupling ratio $r_v$ is $0.91$ for a pure
$\sigma$-coupling $(C_1)$ of 
the nucleon and the $\Delta$, 
while for the case of a
partial $\zeta$-coupling one obtains ${r_v\ge 0.68}$.
In summary it can be seen that the inclusion of the baryon resonances 
may lead to a supersoft equation of state, but since this coupling
cannot be fixed, the EOS cannot be
predicted unambigiously from this chiral approach.\\
Fig.\ref{condensates} shows the behavior of the strange and
non-strange condensates and the resulting baryon masses as a function 
of temperature for vanishing
chemical potential with and without baryon resonances.
The behaviour of the chiral condensates as a function of temperature
depends on the significant contribution of the resonances at
and above the transition temperature.
 In the case that no
resonances are included, one observes a smooth transition to small
expectation values of the condensates, while for the case with included
resonances both scalar fields jump sequentially to lower values.
This is due to the much larger number of
degrees of freedom which accelerate the process of reducing the 
condensates and increasing the scalar density. 
This finally leads to a first
order phase transition. 
In contrast the masses of the vector mesons are
predicted to stay nearly constant, since there is no
direct $\sigma$-$\omega$ coupling term included on the mean-field
level \cite{paper2,paper3}. 
The change of the hadronic masses in the hot and dense medium, as
obtained from chiral arguments, shows that the assumptions of 
vacuum masses and also that of a universal linear drop with density
are at least questionable because of the systems's strong nonlinear
density behaviour. So the determination of 
freeze-out constants ($T, \mu_B, \mu_S$)  
have to be taken cautiously. To demonstrate this, 
we calculate the particle ratios in the chiral model, using 
the values of the freeze-out parameters 
for $S+Au$ collisions at 200 AGeV
as obtained from an ideal gas
model \cite{brau96}.
The results including the feeding from decays of higher resonances 
are shown in figure \ref{ratios}. 
One observes that the change of the masses in the hot 
and dense medium (especially the baryon masses) leads to drastically
altered particle ratios. Further examinations are under way.
\bibliography{preprint}
\bibliographystyle{prsty}
\end{document}